\documentclass[onecolumn]{revtex4}

\usepackage{graphicx}
\usepackage{dcolumn}
\usepackage{bm}
\usepackage{color}
\usepackage{amssymb,amsmath}

\input epsf

\begin{document}

\title{\Large Statefinder Description in Generalized Holographic and Ricci Dark Energy Models}

\author{\bf Piyali Bagchi Khatua$^1$\footnote{piyali.bagchi@yahoo.co.in}
and Ujjal Debnath$^2$\footnote{ujjaldebnath@yahoo.com} }

\affiliation{$^1${Department of Computer Sc and Engg, Netaji Subhas Engineering College, Garia, Kolkata-700 152, India.\\
$^2$Department of Mathematics, Bengal Engineering and Science
University, Shibpur, Howrah-711 103, India.} }

\date{\today}

\begin{abstract}
We have considered the generalized holographic and generalized
Ricci dark energy models for acceleration of the universe. If the
universe filled with only GHDE/GRDE the corresponding deceleration
parameter, EOS parameter and statefinder parameters have been
calculated. Next we have considered that the mixture of GHDE/GRDE
and dark matter in interacting and non-interacting situations.
Also the mixture of GHDE/GRDE and generalized Chaplygin gas have
been analyzed during evolution of the universe. The natures of
above mentioned parameters have been investigated for interacting
and non-interacting situations. Finally, it follows that the
prescribed models derive the acceleration of the universe.
\end{abstract}

\maketitle

\section{\normalsize\bf{Introduction}}

Recent observations of the luminosity of type Ia supernovae
indicate [1 - 3] an accelerated expansion of the universe and lead
to the search for a new type of matter which violates the strong
energy condition. The matter content responsible for such a
condition to be satisfied at a certain stage of evolution of the
universe is referred to as dark energy. The universe is spatially
flat which consists of about 70\% dark energy with negative
pressure, 30\% cold dark matters with baryons and some radiation.
In the observational cosmology of dark energy, the
equation-of-state parameter (EoS) $w=\frac{p}{\rho}$ plays a
central role, where $p$ and $\rho$ are pressure and energy density
of the corresponding field respectively. For accelerating
expansion of the universe, the EoS of dark energy must satisfy $w
< -1/3$. To explain the current accelerated expansion, many models
have been presented, such as cosmological constant, quintessence
[4,5], phantom [6], quintom [7] etc. The simplest candidate of the
dark energy is a very small positive time independent cosmological
constant $\Lambda$, for which $w = -1$. The cosmological constant
remains unchanged while the energy densities of dust matter and
radiation decrease rapidly with the expansion of our universe.
Many dynamical dark energy models have been proposed as
alternatives to the cosmological constant. We will discuss here
two types of dark energy candidates i.e., holographic and Ricci
dark energies. \\

The dark energy problem may be in essence an issue of quantum
gravity [8]. However, by far, we have no any complete theory of
quantum gravity, so it seems that we have to consider the effects
of gravity in some effective quantum field theory in which some
fundamental principles of quantum gravity should be taken into
account. The holographic principle [9] is just a fundamental
principle of quantum gravity, and based on the effective quantum
field theory, it is pointed out that the quantum zero-point energy
of a system with size $L$ should not exceed the mass of a black
hole with the same size [10], i.e. $L^3 \Lambda^4 \leq L M_p^{2}$
where $\Lambda$ is the ultraviolet (UV) cut-off of the effective
quantum field theory, which is closely related to the quantum
zero-point energy density, and $M_p = 1/\sqrt{8\pi G}$ is the
reduced Planck mass. This observation relates the UV cut-off of a
system to its infrared (IR) cutoff. When we take the whole
universe into account, the vacuum energy related to this
holographic principle can be viewed as dark energy. The largest IR
cutoff $L$ is chosen by saturating the inequality, so that we get
the holographic dark energy density [11] $\rho_{\Lambda}=3c^2
M_p^{2}L^{-2}$ where $c$ is a numerical constant characterizing
all of the uncertainties of the theory, whose value can only be
determined by observations. Here $L$ is define by $L=ar(t)$ where
$a$ is a scale factor and $r(t)$ can be obtained from
$\int_{0}^{r(t)}\frac{dr}{\sqrt{1-kr^{2}}}=\int_{t}^{\infty}\frac{dt}{a}=\frac{R_{E}}{a}$,
where $k (=0,\pm 1)$ is the curvature index and $R_{E}$ is the
radius of event horizon [12] and we get
$r(t)=\frac{sin(\sqrt{|k|}R_{E}/a)}{\sqrt{|k|}}$. On the basis of
the holographic principle proposed by [13] several others have
studied holographic model for dark energy [14]. Employment of
Friedman equation [12] $\rho=3M_{p}^{2}H^{2}$ where $\rho$ is the
total energy density and taking $L=H^{-1}$ one can find
$\rho_{m}=3(1-c^{2})M_{p}^{2}H^{2}$. Thus either $\rho_{m}$ or
$\rho_{\Lambda}$ behaves like $H^{2}$. Thus, dark energy results
as pressureless. If we take $L$ as the size of the current
universe, say, the Hubble radius $\frac{1}{H}$ then the dark
energy density will be close to the observational result. But,
neither dark energy, nor dark matter has laboratory evidence
for its existence directly.\\

Many candidates such as cosmological constant, quintessence,
phantom, holographic dark energy, etc. have been proposed to
explain the acceleration. Ricci dark energy, which is a kind of
holographic dark energy [15] taking the square root of the inverse
Ricci scalar as its infrared cutoff and this model is also
phenomenologically viable. Gao et al [16] proposed the dark energy
density proportional to the Ricci scalar $R$ i.e.,
$\rho_{X}\propto R$ is called the {\it Ricci dark energy}. This
model works fairly well in fitting the observational data, and it
could also help to understand the coincidence problem. Moreover,
in this model the presence of event horizon is not presumed, so
the causality problem is avoided. Assuming the black hole is
formed by gravitation collapsing of the perturbation in the
universe, the maximal black hole can be formed is determined by
the casual connection scale $R_C$ [17], given by $R_C =
1/\sqrt{Max(\dot{H} +2H^2,- \dot{H})}$ for a flat universe, where
$H$ is the Hubble parameter, and if $ R_C = 1/\sqrt{(\dot{H}
+2H^2)}$, it could be consistent with the current cosmological
observations when the vacuum density appears as an independently
conserved energy component. As we know, in flat FRW universe, the
Ricci scalar is $R =- 6(\dot{H} + 2H^2)$, which means the $R_C$ is
proportional to $R$ and if one choices the casual connection scale
$R_C$ as the IR cutoff, the Ricci dark energy model is also
obtained. There are several works on this Ricci dark energy model [18]. \\

In this work, we have first defined the generalized holographic
and generalized Ricci dark energy models in section II. The
statefinder parameters in two fluid system has been presented in
section III. The equation of state and statefinder parameters for
generalized holographic and generalized Ricci dark energy models
without dark matter and with dark matter in non-interacting and
interacting scenarios have been calculated in sections IV - VI. In
sections VII and VIII, the equation of state and statefinder
parameters for generalized holographic and generalized Ricci dark
energy models with generalized Chaplygin gas in non-interacting
and interacting scenarios have been studied. Finally, some
fruitful conclusions have been drawn.\\

\section{\normalsize\bf{Two Dark Energy Models : Definitions}}

Xu et al [19] proposed two types of dark energy models, i.e.,
generalized holographic and generalized Ricci dark energy models
as follows:\\

(i) {\bf{Generalized Holographic dark energy (GHDE):}} The energy
density of GHDE is given by,

\begin{equation}
\rho_{h}=3c^2m_{p}^2H^2f(R/H^2)
\end{equation}

where $c$ is a numerical constant and $f(x)$ is a positive
function defined as, $f(x)=\alpha x+(1-\alpha)$, $\alpha$ is a
constant. When $\alpha=0$ then
$f(x)=1$ and when $\alpha=1$ then $f(x)=x$. \\

(ii) {\bf{Generalized Ricci dark energy (GRDE):}} The energy
density of GRDE is given by,

\begin{equation}
\rho_{r}=3c^2m_{p}^2R~g(H^2/R)
\end{equation}

where  $g(y)$ is a positive function defined as, $g(y)=\beta
y+(1-\beta)$, $\beta$ is a constant. When $\beta=0$
then $g(y)=1$ and when $\beta=1$ then $g(y)=y$.\\

It is interesting that when $f(x)=g(y)=1$, we recover the energy
densities of original holographic and Ricci dark energies. Also
when $f(x)=x$ and $g(y)=y$, we recover interchangeably the energy
densities of original holographic and Ricci dark energies. Also
when $\alpha=0$ and $\beta=1$ or when $\alpha=1$ and $\beta=0$
then $\rho_{h}$ and $\rho_{r}$ both are same. Now comparing (1)
and (2), we see that when $\beta=1-\alpha$ the generalized Ricci
dark energy reduces to the generalized holographic dark energy and
vice versa. \\

The Ricci scalar for non-flat FRW universe is given by,
$R=-6(\dot{H}+2H^2+\frac{k}{a^2})$, where $k$ = +1, 0 and $-1$ for
closed, flat, and open geometries, respectively. We shall consider
the flat universe ($k=0$) where,

\begin{equation}
R=-6(\dot{H}+2H^2)
\end{equation}

In the next section, we shall prescribe the statefinder diagnostic
pair and deceleration parameter in two fluid system.\\

\section{\normalsize\bf{Statefinder diagnostics for two fluid systems}}

The Einstein field equations for the mixture of dark matter and
dark energy are,

\begin{equation}
H^{2}=\frac{1}{3}(\rho_{m}+\rho_{X})
\end{equation}
and
\begin{equation}
\dot{H}=-\frac{1}{2}(\rho_{m}+\rho_{X}+p_{m}+p_{X})
\end{equation}

Also the conservation equation is given by

\begin{equation}
\dot{\rho}_{m}+\dot{\rho}_{X}+3H(\rho_{m}+\rho_{X}+p_{m}+p_{X})=0
\end{equation}

where $p_{X}$ and $p_{m}$ denote pressures and $\rho_{X}$ and
$\rho_{m}$ are the energy densities of dark matter and dark energy
respectively. Thus here $p_{X}=p_{h},~\rho_{X}=\rho_{h}$ for
generalized holographic dark energy and
$p_{X}=p_{r},~\rho_{X}=\rho_{r}$ for
generalized Ricci dark energy respectively.\\

The flat Friedmann model which is analyzed in terms of the
statefinder parameters [20]. The trajectories in the $\{r, s\}$
plane of different cosmological models shows different behavior.
The statefinder diagnostic of SNAP observations used to
discriminate between different dark energy models. The statefinder
diagnostic pair is constructed from the scale factor $a(t)$. The
statefinder diagnostic pair is denoted as $\{r,s\}$ and defined
as,

\begin{equation}
 r=\frac{\dddot{a}}{a H^3}
 ~~~\text{and}~~~
s=\frac{r-1}{3(q-\frac{1}{2})}
\end{equation}

where $q$ is the deceleration parameter given by, $q=-\frac{a
{\ddot{a}}}{{\dot{a}}^2}$.\\

The parameters can be expressed as

\begin{equation}
 r=1+\frac{9}{2(\rho_{X}+\rho_{m})}\left(\frac{\partial p_{X}}{\partial
 \rho_{X}}(\rho_{X}+p_{X})+\frac{\partial p_{m}}{\partial
 \rho_{m}}(\rho_{m}+p_{m})\right)
 \end{equation}

\begin{equation}
s=\frac{1}{(p_{X}+p_{m})}\left(\frac{\partial p_{X}}{\partial
 \rho_{X}}(\rho_{X}+p_{X})+\frac{\partial p_{m}}{\partial
 \rho_{m}}(\rho_{m}+p_{m})\right)
\end{equation}

and

\begin{equation}
q=\frac{1}{2}+\frac{3}{2}\left(\frac{p_{X}+p_{m}}{\rho_{X}+\rho_{m}}\right)
\end{equation}

\section{\normalsize\bf{GHDE and GRDE models without Dark
Matter}}

We follow the work of Kim et al [21] in generalized holographic
and generalized Ricci dark energy models. Here we consider that
the universe is filled with only generalized holographic/Ricci
dark energy. In this case $\rho_{m}=0=p_{m}$, so the first
Friedmann equation (4) can be written as,

\begin{equation}
H^2 = \frac{1}{3}\rho_{h}
\end{equation}

Combining (1) and (3) we get the GHDE density as,

\begin{equation}
\rho_{h}=3c^2[-6\alpha\dot{H}+(1-13\alpha)H^2]
\end{equation}

Combining this equation with (11) we get,

\begin{equation}
6\alpha\dot{H}=(1-13\alpha-\frac{1}{c^2})H^2
\end{equation}

Solving the above differential equation we get,

\begin{equation}
H^2=H_{0}^{2} a^{\frac{1}{3\alpha}(1-13\alpha-\frac{1}{c^2})}
\end{equation}

where $H_{0}$ is the integrating constant which is the present
value of $H$. So from (11) and (14) we have,

\begin{equation}
\rho_{h}=3H^2=3H_{0}^{2}
a^{\frac{1}{3\alpha}(1-13\alpha-\frac{1}{c^2})}
\end{equation}

Second Friedmann equation is given by (from (5)),

\begin{equation}
\dot{H}=-\frac{1}{2}(p_{h}+\rho_{h})
\end{equation}

Using (13) - (15), the equation (16) becomes,

\begin{equation}
p_{h}=\frac{1}{3\alpha}(4\alpha+\frac{1}{c^2}-1)H_{0}^{2}
a^{\frac{1}{3\alpha}(1-13\alpha-\frac{1}{c^2})}
\end{equation}

So the equation of state $w_{h}$ for generalized holographic dark
energy model is defined as,

\begin{equation}
w_{h}=\frac{p_{h}}{\rho_{h}}=\frac{1}{9\alpha c^{2}}\left(4\alpha
c^{2}-c^2+1\right)
\end{equation}

This model generates dark energy if $w_{h}<-1/3$ ~i.e., if
$\alpha<\frac{c^{2}-1}{7c^{2}}$. The equation of state parameter
$w_{h}$ has been drawn against $\alpha$ and $c$ in fig.1.
From the figure, we have seen that $w_{h}$ decreases from positive to negative
as $\alpha$ decreases and $c$ increases.\\

\begin{figure}[h!]
\includegraphics[height=2.5in]{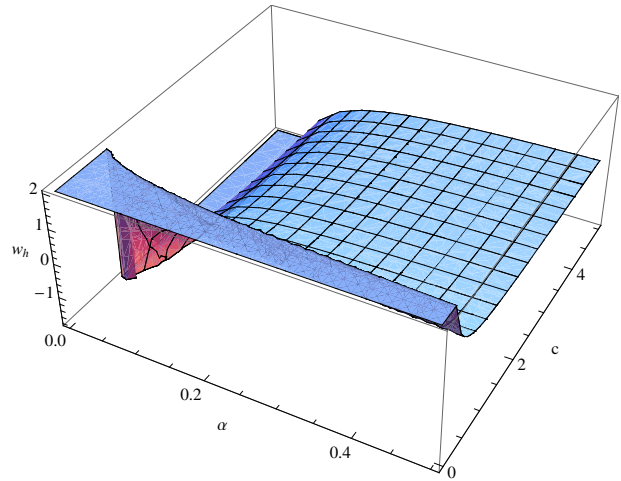}~~~~ \caption{The
variation of $w_h$ against $\alpha$ and $c$.}
\end{figure}

\begin{figure}[h!]
\includegraphics[height=2.5 in]{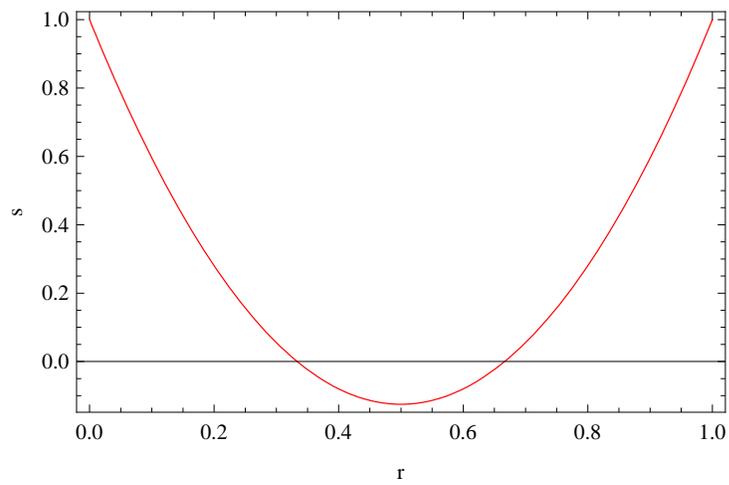}~~~~ \caption{The
variation of $r$ against $s$.}
\end{figure}

\begin{figure}[h!]
\includegraphics[height=2.5in]{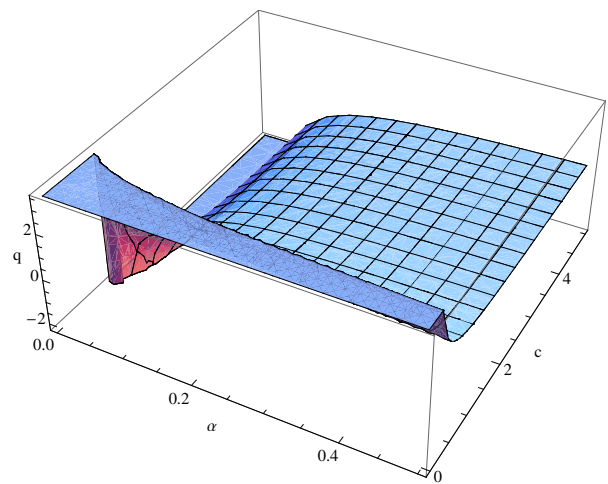}~~~~ \caption{The
variation of $q$ against $\alpha$ and $c$.}
\end{figure}

Now from (15) and (17) we get,

\begin{equation}
\frac{\partial p_{h}}{\partial
\rho_{h}}=\frac{\partial{p_{h}}/\partial
a}{\partial{\rho_{h}}/\partial a} =\frac{1}{9\alpha
c^{2}}\left(4\alpha c^{2}-c^2+1\right)
\end{equation}

So from (8), (9), (10), (18) and (19) we obtain $r$, $s$ and $q$
as
\begin{equation}
r=1+\frac{1}{18\alpha^{2}c^{4}}\left(13\alpha
c^{2}-c^{2}\right)\left(4\alpha c^{2}-c^{2}+1\right)
\end{equation}

\begin{equation}
s=\frac{1}{9\alpha c^{2}}\left(13\alpha c^{2}-c^2+1\right)
\end{equation}

and

\begin{equation}
q=\frac{1}{6\alpha c^{2}}\left(7\alpha c^{2}-c^2+1\right)
\end{equation}

The universe will be accelerating if $q<0$ i.e., if
$\alpha<\frac{c^{2}-1}{7c^{2}}$.\\

The variation of $r$ against $s$ has been drawn in fig.2. From the
figure, we have seen that when $r$ increases, $s$ first decreases
from positive to negative upto about $r=0.5$ and then $s$
increases from negative to positive. The deceleration parameter
$q$ has been drawn against $\alpha$ and $c$ in fig.3. From the
figure, we have seen that $q$ decreases
from positive to negative as $\alpha$ decreases and $c$ increases.\\

If the universe is filled with generalized Ricci dark energy
rather than generalized holographic dark energy then all the above
solutions are valid provided $\alpha=1-\beta$. So the universe
will be accelerating if $\beta>\frac{8c^{2}-1}{7c^{2}}$.\\

\section{\normalsize\bf{GHDE and GRDE models with Dark
Matter : Non-Interacting Scenario}}

Here we consider the universe is filled with the mixture of dark
matter and GHDE and also there is no interaction between them. In
this case the first Friedmann equation (4) can be written as,

\begin{equation}
H^2 = \frac{1}{3}\left(\rho_{h}+\rho_{m}\right)
\end{equation}

Since there is no interaction, so dark matter and GHDE are
separately conserved and hence the energy conservation equations
for dark matter and dark energy are (from (6)),

\begin{equation}
\dot\rho_{m}+3H(\rho_{m}+p_{m})=0
\end{equation}

and

\begin{equation}
\dot\rho_{h}+3H(\rho_{h}+p_{h})=0
\end{equation}

Assume that the equation of state for dark matter is
$p_{m}=w_{m}\rho_{m}$. Putting it in the above equation and after
solving the differential equation we get,

\begin{equation}
\rho_{m}=\rho_{m0} a^{-3(1+w_{m})}
\end{equation}

where $\rho_{m0}$ is the integrating constant. Combining (12),
(23) and (26) we have,

\begin{equation}
6\alpha c^{2}\dot{H}+(13\alpha
c^{2}-c^{2}+1)H^2=\frac{1}{3}\rho_{m0}a^{-3(1+w_{m})}
\end{equation}

and after solving we get,

\begin{equation}
H^2=\frac{\rho_{m0}a^{-3(1+w_{m})}}{3\{(4\alpha
c^{2}-c^{2}+1)-9\alpha c^{2}w_{m} \}}+H_{1}^{2}a^{-\frac{(13\alpha
c^{2}-c^{2}+1)}{3\alpha c^{2}}}
\end{equation}

where $H_{1}$ is the integrating constant. From (23), (25), (26)
and (28) we have,

\begin{equation}
\rho_{h}=3H^2-\rho_{m}=\left[\frac{1}{\{(4\alpha
c^{2}-c^{2}+1)-9\alpha c^{2}w_{m}
\}}-1\right]~\rho_{m0}a^{-3(1+w_{m})}+3H_{1}^{2}a^{-\frac{(13\alpha
c^{2}-c^{2}+1)}{3\alpha c^{2}}}
\end{equation}
and
\begin{equation}
p_{h}=\left[\frac{1}{\{(4\alpha c^{2}-c^{2}+1)-9\alpha c^{2}w_{m}
\}}-1\right]~w_{m}\rho_{m0}a^{-3(1+w_{m})}+\frac{(4\alpha
c^{2}-c^{2}+1)}{3\alpha c^{2}}H_{1}^{2}a^{-\frac{(13\alpha
c^{2}-c^{2}+1)}{3\alpha c^{2}}}
\end{equation}

\begin{figure}[!h]

\includegraphics[height=2.5in]{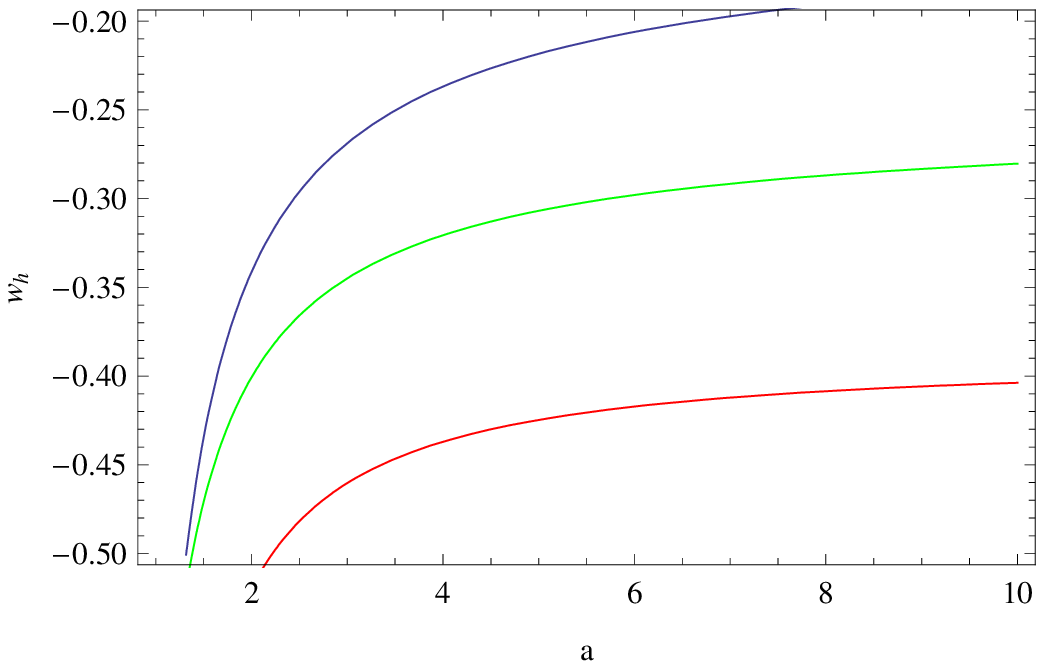}~\\
\vspace{1mm}\caption{The variation of $w_{h}$ against $a$ for
$w_{m}=.01, \rho_{m0}=1, H_{1}=1, c=2 $ for different values of
$\alpha=.1,.12,.15$.}

\vspace{6mm}

\end{figure}

\begin{figure}[!h]
\includegraphics[height=2.5in]{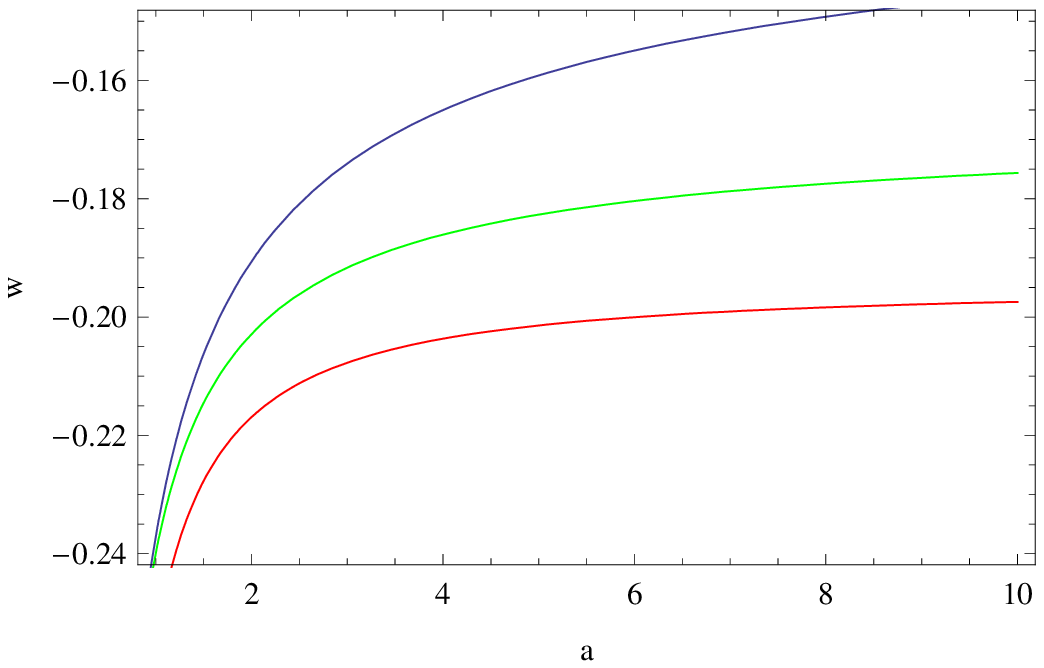}~\\
\vspace{1mm}\caption{The variation of $w$ against $a$ for
$w_{m}=.01, \rho_{m0}=1, H_{1}=1, c=2 $ for different values of
$\alpha=.1,.12,.15$.}

\vspace{6mm}

\end{figure}

So the equation of state $w_{h}$ for GHDE in this non-interacting
scenario is obtained as,

\begin{equation}
w_{h}=\frac{p_{h}}{\rho_{h}}=\frac{\left[\frac{1}{\{(4\alpha
c^{2}-c^{2}+1)-9\alpha c^{2}w_{m}
\}}-1\right]~w_{m}\rho_{m0}a^{-3(1+w_{m})}+\frac{(4\alpha
c^{2}-c^{2}+1)}{3\alpha c^{2}}H_{1}^{2}a^{-\frac{(13\alpha
c^{2}-c^{2}+1)}{3\alpha c^{2}}} }{\left[\frac{1}{\{(4\alpha
c^{2}-c^{2}+1)-9\alpha c^{2}w_{m}
\}}-1\right]~\rho_{m0}a^{-3(1+w_{m})}+3H_{1}^{2}a^{-\frac{(13\alpha
c^{2}-c^{2}+1)}{3\alpha c^{2}}} }
\end{equation}

\begin{figure}[!h]
\includegraphics[height=2.0in]{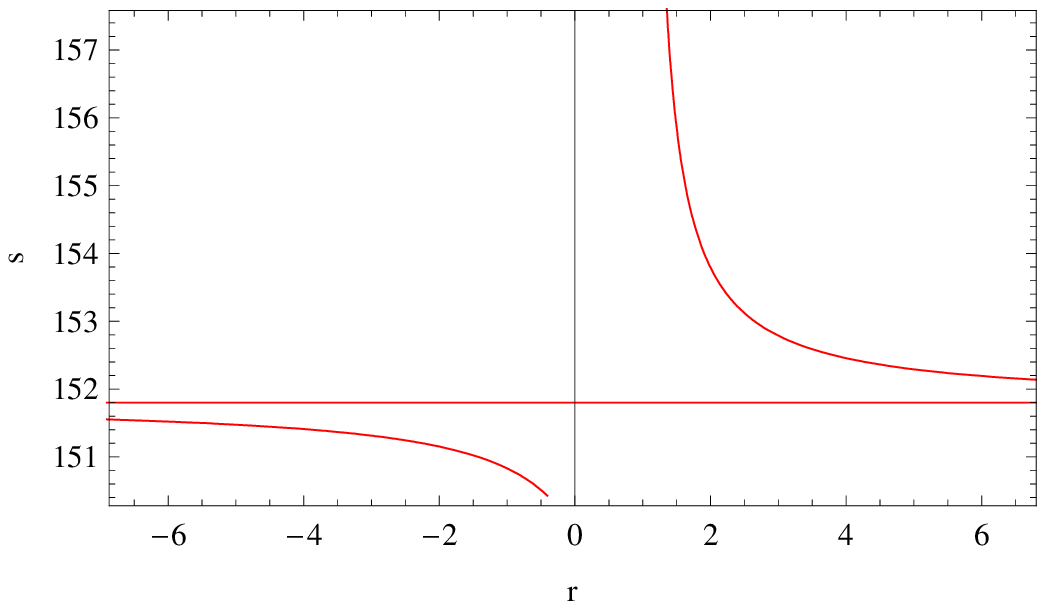}~~~~ \caption{The
variation of $r$ against $s$ for $w_{m}=.01, \rho_{m0}=1, H_{1}=1,
c=2 $ and $\alpha=.1$.} \vspace{7mm}
\end{figure}

\begin{figure}[!h]
\includegraphics[height=2.5in]{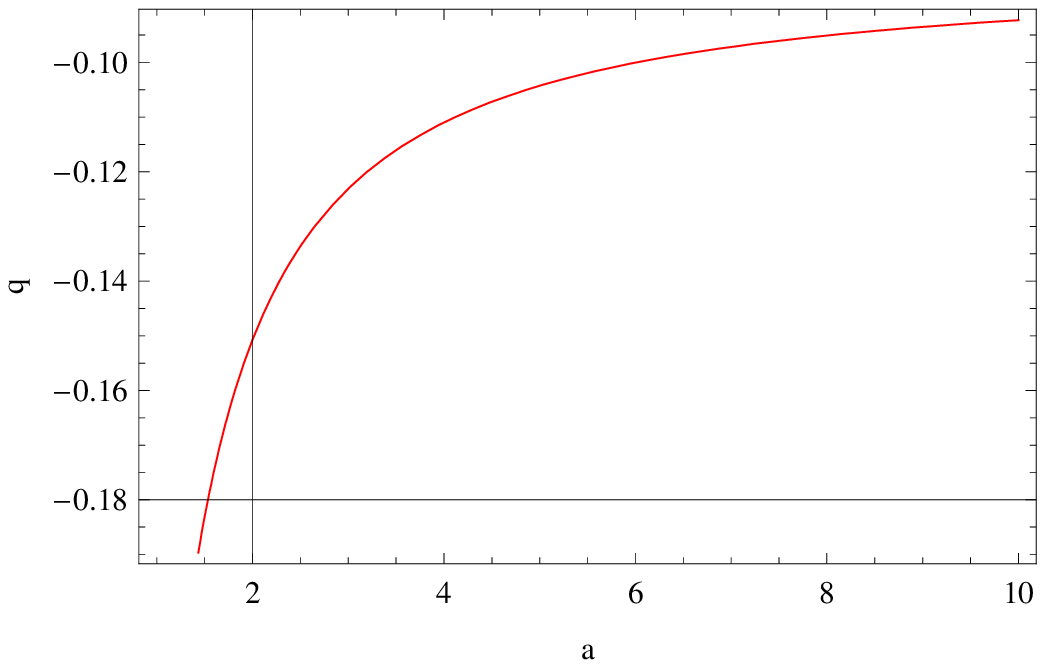}~~~~ \caption{The
variation of $q$ against $a$ for $w_{m}=.01, \rho_{m0}=1, H_{1}=1,
c=2 $ and $\alpha=.1$.} \vspace{7mm}
\end{figure}

Also the equation state for combined fluid is obtained as

\begin{equation}
w=\frac{p_{m}+p_{h}}{\rho_{m}+\rho_{h}}=\frac{\frac{w_{m}\rho_{m0}a^{-3(1+w_{m})}}{\{(4\alpha
c^{2}-c^{2}+1)-9\alpha c^{2}w_{m} \}}+\frac{(4\alpha
c^{2}-c^{2}+1)}{3\alpha c^{2}}H_{1}^{2}a^{-\frac{(13\alpha
c^{2}-c^{2}+1)}{3\alpha c^{2}}}
}{\frac{\rho_{m0}a^{-3(1+w_{m})}}{\{(4\alpha
c^{2}-c^{2}+1)-9\alpha c^{2}w_{m}
\}}+3H_{1}^{2}a^{-\frac{(13\alpha c^{2}-c^{2}+1)}{3\alpha c^{2}}}
}
\end{equation}

The above expressions are very complicated, so the variation of
$w_{h}$ against $a$ has been drawn in fig.4 for
 $w_{m}=.01, \rho_{m0}=1, H_{1}=1, c=2 $ with different
values of $\alpha=.1,.12,.15$. Fig.5 ~represents the variation of
$w$ against $a$ for $w_{m}=.01, \rho_{m0}=1, H_{1}=1, c=2 $ for
different values of $\alpha=.1,.12,.15$. Also from (8), (9), (10),
(26), (29) and (30) we get graphs of $s$ against $r$ and $q$
against $a$ in figures 6 and 7 respectively. We have seen that $s$
first increases from some positive value to $+\infty$ and after
that $s$ also increases from $-\infty$ to some positive value as
$r$ increases from negative to positive. We have seen that the EOS
for GHDE $w_{h}$ keeps negative sign as evolution of the universe.
When we have considered GHDE with dark matter without interaction,
the EOS for combined fluid generates the negative sign. The
deceleration parameter also gives us the negative sign, so the
non-interacting model also generates dark energy. When $\alpha$
is replaced by $(1-\beta)$, we get the similar results for GRDE in with dark matter.\\

\section{\normalsize\bf{GHDE and GRDE models with Dark
Matter : Interacting Scenario}}

Here we consider the universe is filled with the mixture of dark
matter and GHDE and also there is an interaction between them. So
dark matter and GHDE are not separately conserved and let the
interaction term is defined by $3\delta H\rho_{m}$, where $\delta$
is the interaction parameter. From (6), we get the relations

\begin{equation}
\dot\rho_{m}+3H(\rho_{m}+p_{m})=-3\delta H\rho_{m}
\end{equation}

and

\begin{equation}
\dot\rho_{h}+3H(\rho_{h}+p_{h})=3\delta H\rho_{m}
\end{equation}

Assume that the equation of state for dark matter is
$p_{m}=w_{m}\rho_{m}$. Putting it in the above equation and after
solving the differential equation we get,

\begin{equation}
\rho_{m}=\rho_{m1} a^{-3(1+w_{m}+\delta)}
\end{equation}

where $\rho_{m1}$ is the integrating constant. Combining (12),
(23) and (35) we have,

\begin{equation}
6\alpha c^{2}\dot{H}+(13\alpha
c^{2}-c^{2}+1)H^2=\frac{1}{3}\rho_{m1}a^{-3(1+w_{m}+\delta)}
\end{equation}

and after solving we get,

\begin{equation}
H^2=\frac{\rho_{m1}a^{-3(1+w_{m}+\delta)}}{3\{(4\alpha
c^{2}-c^{2}+1)-9\alpha c^{2}(w_{m}+\delta)
\}}+H_{2}^{2}a^{-\frac{(13\alpha c^{2}-c^{2}+1)}{3\alpha c^{2}}}
\end{equation}

where $H_{2}$ is the integrating constant. From (23), (34), (35)
and (37) we have,

\begin{equation}
\rho_{h}=\left[\frac{1}{\{(4\alpha c^{2}-c^{2}+1)-9\alpha
c^{2}(w_{m}+\delta)
\}}-1\right]~\rho_{m0}a^{-3(1+w_{m}+\delta)}+3H_{2}^{2}a^{-\frac{(13\alpha
c^{2}-c^{2}+1)}{3\alpha c^{2}}}
\end{equation}
and
\begin{equation}
p_{h}=\left[\frac{w_{m}+\delta}{\{(4\alpha c^{2}-c^{2}+1)-9\alpha
c^{2}(w_{m}+\delta)
\}}-w_{m}\right]~\rho_{m1}a^{-3(1+w_{m}+\delta)}+\frac{(4\alpha
c^{2}-c^{2}+1)}{3\alpha c^{2}}H_{2}^{2}a^{-\frac{(13\alpha
c^{2}-c^{2}+1)}{3\alpha c^{2}}}
\end{equation}

\begin{figure}[!h]
\includegraphics[height=2.5in]{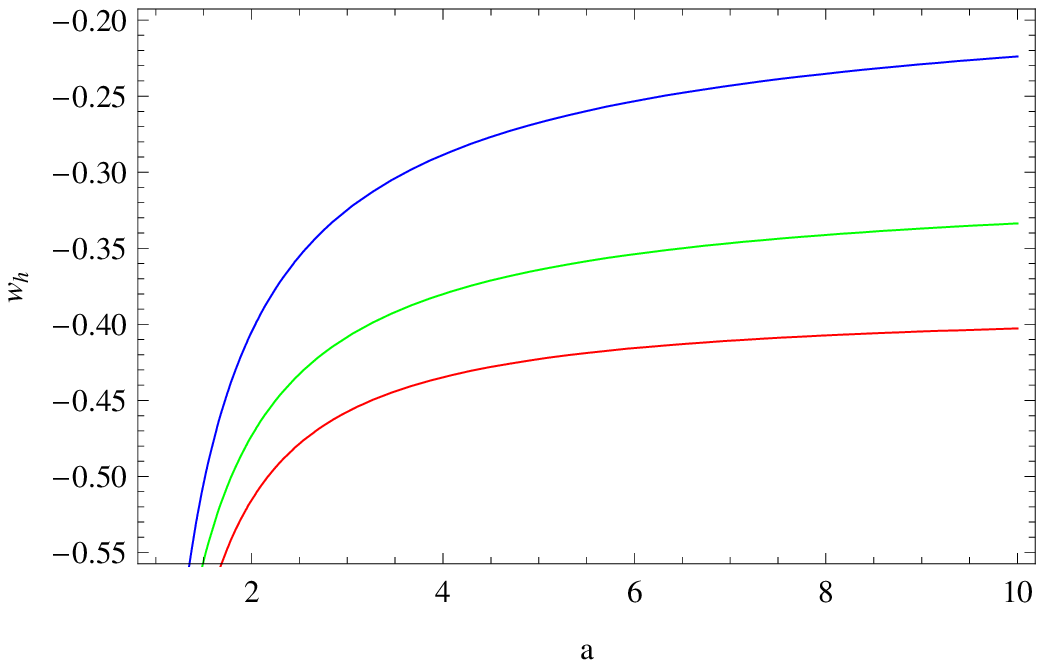}~\\
\vspace{1mm}\caption{The variation of $w_{h}$ against $a$ for
$w_{m}=.01, \rho_{m1}=1, H_{2}=1, c=2, \delta=.01 $ for different
values of $\alpha=.1,.12,.15$.}

\vspace{2mm}

\end{figure}

\begin{figure}[!h]
\includegraphics[height=2.5in]{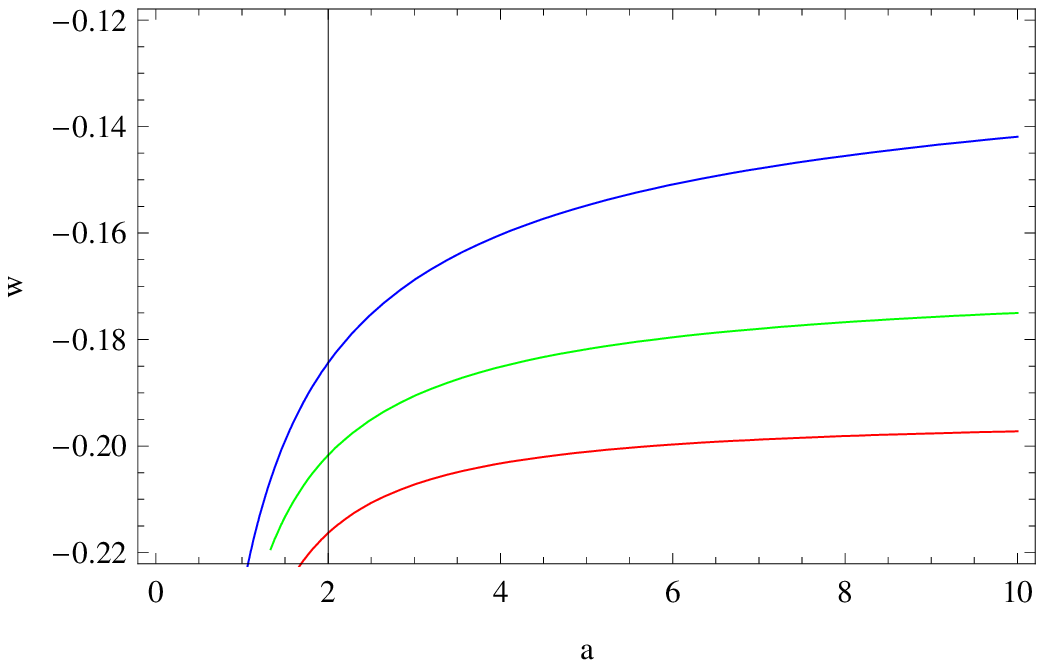}~\\
\vspace{1mm}\caption{The variation of $w$ against $a$ for
$w_{m}=.01, \rho_{m1}=1, H_{2}=1, c=2, \delta=.01 $ for different
values of $\alpha=.1,.12,.15$.}

\end{figure}

\begin{figure}[!h]
\includegraphics[height=2.5in]{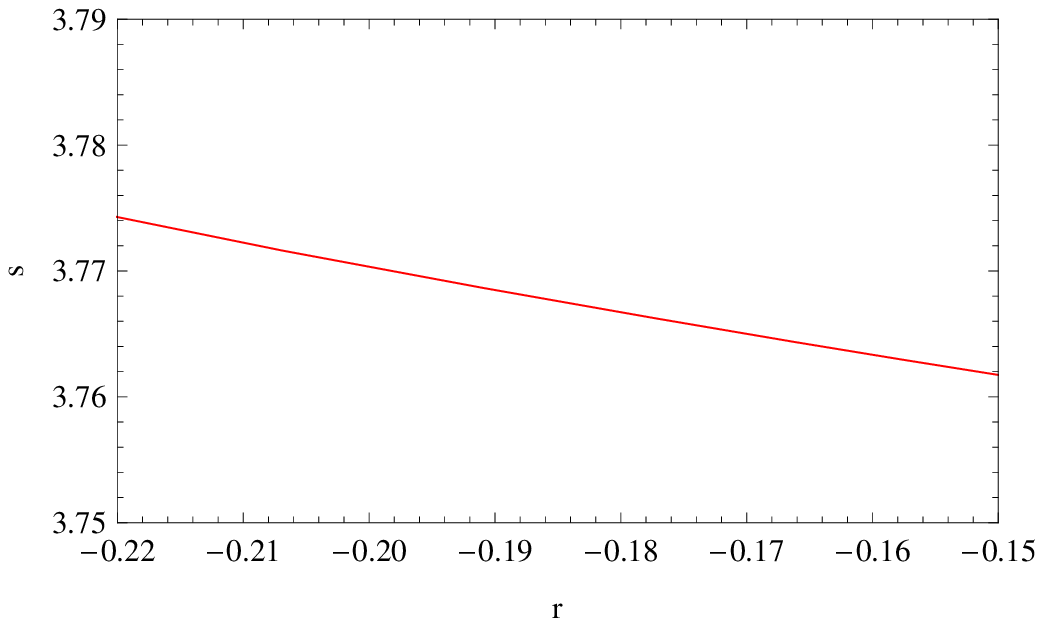}~\\
\vspace{1mm}\caption{The variation of $s$ against $r$ for
$w_{m}=.01, \rho_{m1}=1, H_{2}=1, c=2, \delta=.01, \alpha=.1$.}

\end{figure}

\begin{figure}[!h]
\includegraphics[height=2.5in]{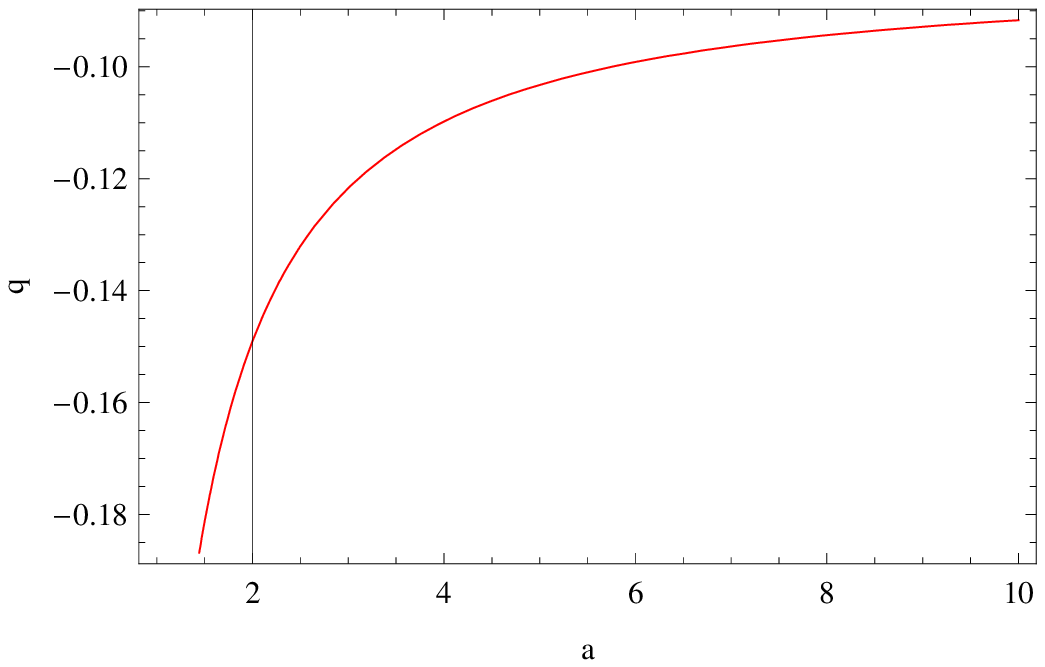}~\\
\vspace{1mm}\caption{The variation of $q$ against $a$ for
$w_{m}=.01, \rho_{m1}=1, H_{2}=1, c=2, \delta=.01, \alpha=.1$.}

\end{figure}

So the equation of state for GHDE $w_{h}$ is obtained as,

\begin{equation}
w_{h}=\frac{\left[\frac{w_{m}+\delta}{\{(4\alpha
c^{2}-c^{2}+1)-9\alpha c^{2}(w_{m}+\delta)
\}}-w_{m}\right]~\rho_{m1}a^{-3(1+w_{m}+\delta)}+\frac{(4\alpha
c^{2}-c^{2}+1)}{3\alpha c^{2}}H_{2}^{2}a^{-\frac{(13\alpha
c^{2}-c^{2}+1)}{3\alpha c^{2}}} }{\left[\frac{1}{\{(4\alpha
c^{2}-c^{2}+1)-9\alpha c^{2}(w_{m}+\delta)
\}}-1\right]~\rho_{m1}a^{-3(1+w_{m}+\delta)}+3H_{2}^{2}a^{-\frac{(13\alpha
c^{2}-c^{2}+1)}{3\alpha c^{2}}} }
\end{equation}

Also the equation state for combined fluid is obtained as,

\begin{equation}
w=\frac{p_{m}+p_{h}}{\rho_{m}+\rho_{h}}=\frac{\frac{(w_{m}+\delta)\rho_{m1}a^{-3(1+w_{m}+\delta)}}{\{(4\alpha
c^{2}-c^{2}+1)-9\alpha c^{2}(w_{m}+\delta) \}}+\frac{(4\alpha
c^{2}-c^{2}+1)}{3\alpha c^{2}}H_{2}^{2}a^{-\frac{(13\alpha
c^{2}-c^{2}+1)}{3\alpha c^{2}}}
}{\frac{\rho_{m1}a^{-3(1+w_{m}+\delta)}}{\{(4\alpha
c^{2}-c^{2}+1)-9\alpha c^{2}(w_{m}+\delta)
\}}+3H_{2}^{2}a^{-\frac{(13\alpha c^{2}-c^{2}+1)}{3\alpha c^{2}}}
}
\end{equation}

The above expressions are very completed, so the equation of state
parameters $w_{h}$ and $w$ against $a$ have been drawn in figure 8
and 9 respectively for  $w_{m}=.01, \rho_{m1}=1, H_{2}=1, c=2,
\delta=.01$ with different values of $\alpha=.1,.12,.15$. Also
from (8), (9), (10), (35), (38) and (39) we draw the graphs of $s$
against $r$ and $q$ against $a$ in figures 10 and 11 respectively
for  $w_{m}=.01, \rho_{m1}=1, H_{2}=1, c=2, \delta=.01,
\alpha=.1$. We have seen that $s$ always decreases with positive
sign as $r$ increases with negative level. We have seen that the
EOS for GHDE $w_{h}$ keeps negative sign as evolution of the universe.
When we have considered interacting GHDE with dark matter, the EOS for
combined fluid generates the negative sign. The deceleration parameter also gives us
the negative sign, so the interacting model also generates dark energy. When $\alpha$
is replaced by $(1-\beta)$, we get the similar results for GRDE in interaction with dark matter.\\

\section{\normalsize\bf{GHDE and GRDE models with Generalized Chaplygin Gas : Non-Interacting Scenario}}

Let the universe is filled with the mixture of generalized
Chaplygin Gas and GHDE. This generalized Chaplygin Gas is
considered a perfect fluid which follows the adiabatic equation of
state. The equation of Generalized Chaplygin Gas [22] is given by,
\begin{equation}
p_{c}=-B/{\rho}_{c}^{\gamma}~~~~~~~~~~~\text~~~~~~~~~~~~ 0\le
\gamma \le 1, B>0.
\end{equation}
In this case we consider the universe is filled with the mixture
of Chaplygin Gas and GHDE and also
there is no interaction between them.\\

In this case the first Friedmann equation can be written as,

\begin{equation}
H^2 = \frac{1}{3}\left(\rho_{h}+\rho_{c}\right)
\end{equation}

Since there is no interaction, so Chaplygin Gas and GHDE are
separately conserved and hence the energy conservation equations
for Chaplygin Gas and dark energy are,

\begin{equation}
\dot\rho_{c}+3H(\rho_{c}+p_{c})=0
\end{equation}

and

\begin{equation}
\dot\rho_{h}+3H(\rho_{h}+p_{h})=0
\end{equation}

from which we get,
\begin{equation}
\rho_{c}=[B+\rho_{c0}a^{-3(1+\gamma)}]^\frac{1}{1+\gamma}
\end{equation}

where $\rho_{c0}$ is the integrating constant. Hence we get,
\begin{equation}
p_{c}=-[B+\rho_{c0}a^{-3(1+\gamma)}]^\frac{1}{1+\gamma}
+\rho_{c0}a^{-3(1+\gamma)}[B+\rho_{c0}a^{-3(1+\gamma)}]^\frac{-\gamma}{1+\gamma}
\end{equation}

Putting the value of $p_c$ and $\rho_{c}$ we get,
\begin{equation}
6\alpha c^{2}\dot{H}+(13\alpha
c^{2}-c^{2}+1)H^2=\frac{1}{3}[B+\rho_{c0}a^{-3(1+\gamma)}]^\frac{1}{1+\gamma}
\end{equation}

and after solving we get,

\begin{equation}
H^2=\frac{1}{9\alpha c^2}e^{-\frac{(13\alpha c^2-c^2+1)}{3\alpha
c^2}x}\int\left[B+\rho_{c0}e^{-3(1+\gamma)x}\right]^{\frac{1}{1+\gamma}}e^{\frac{(13\alpha
c^2-c^2+1)}{3\alpha c^2}x}dx+H_{c0}^{2}a^{-\frac{(13\alpha
c^{2}-c^{2}+1)}{3\alpha c^{2}}}
\end{equation}

where $x=\ln a$ and $H_{c0}$ is the integrating constant. Hence we
have,

\begin{equation}
\rho_{h}=\frac{1}{3\alpha c^2}e^{-\frac{(13\alpha
c^2-c^2+1)}{3\alpha c^2}x}
~I(x)-\left[B+\rho_{c0}a^{-3(1+\gamma)}\right]^{\frac{1}{1+\gamma}}+3H_{c0}^{2}a^{-\frac{(13\alpha
c^{2}-c^{2}+1)}{3\alpha c^{2}}}
\end{equation}
where,
\begin{equation}
I(x)=\int\left[B+\rho_{c0}e^{-3(1+\gamma)x}\right]^{\frac{1}{1+\gamma}}e^{\frac{(13\alpha
c^2-c^2+1)}{3\alpha c^2}x}dx
\end{equation}

Now, from energy conservation equation we get,
\begin{equation}
p_{h}=-\rho_{h}-\frac{1}{3}.\frac{\partial\rho_{h}}{\partial x}
\end{equation}

which gives,
\begin{eqnarray*}
p_{h}=\frac{(4\alpha c^2 - c^2+1)}{27\alpha^2
c^4}e^{-\frac{(13\alpha c^2-c^2+1)}{3\alpha
c^2}x}\int\left[B+\rho_{c0}e^{-3(1+\gamma)x}\right]^{\frac{1}{1+\gamma}}e^{\frac{(13\alpha
c^2-c^2+1)}{3\alpha c^2}x}dx
\end{eqnarray*}
\begin{equation}
-\frac{1}{9\alpha
c^2}\left[B+\rho_{c0}a^{-3(1+\gamma)}\right]^{\frac{1}{1+\gamma}}+
B\left[B+\rho_{c0}a^{-3(1+\gamma)}\right]^{\frac{-\gamma}{1+\gamma}}+\frac{(4\alpha
c^{2}-c^{2}+1)}{3\alpha c^{2}}H_{c0}^{2}a^{-\frac{(13\alpha
c^{2}-c^{2}+1)}{3\alpha c^{2}}}
\end{equation}

\begin{figure}[!h]
\includegraphics[height=2.5in]{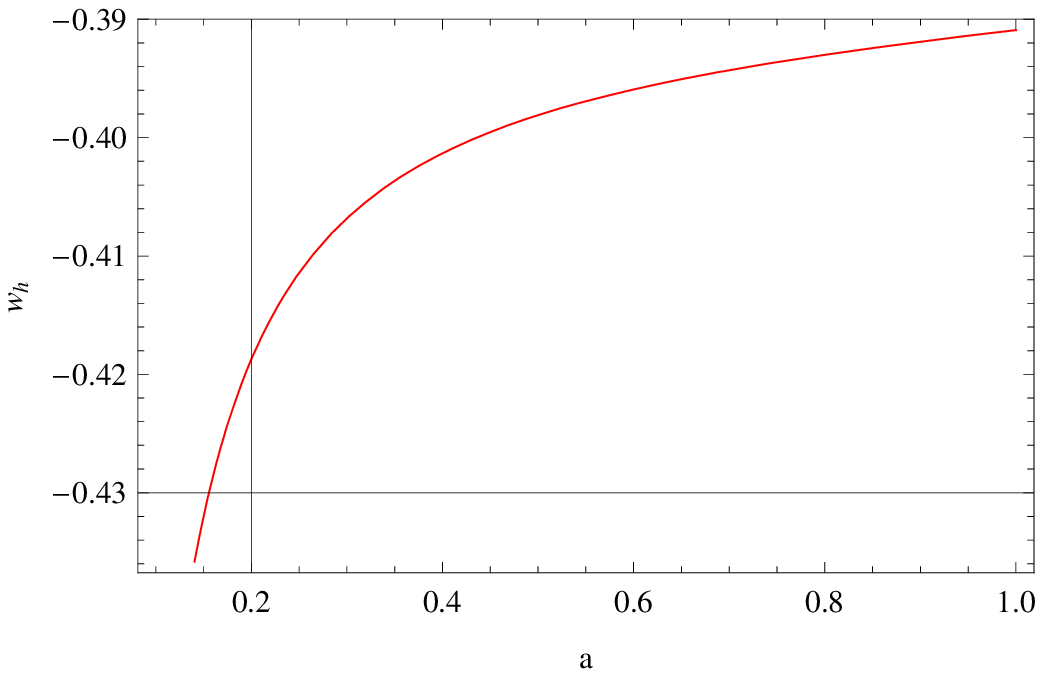}~\\
\vspace{1mm}\caption{The variation of $w_{h}$ against $a$ for $
\rho_{c0}=1, H_{c0}=1, c=2, \alpha=.1, B=1, \gamma=.1$.}

\end{figure}

\begin{figure}[!h]
\includegraphics[height=2.5in]{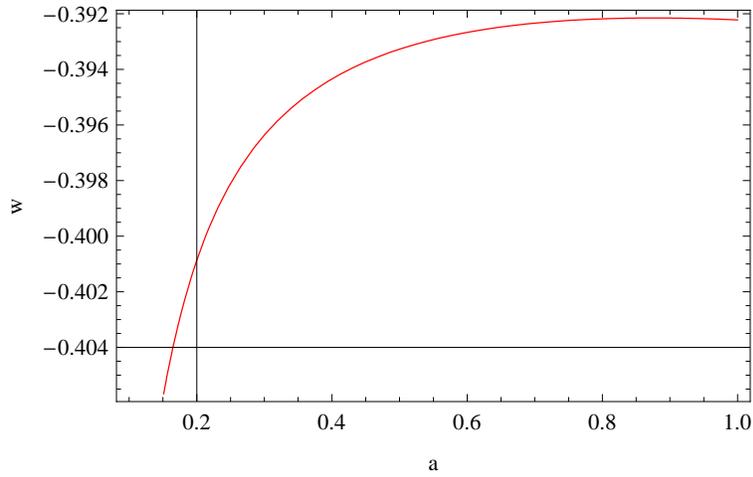}~\\
\vspace{1mm}\caption{The variation of $w$ against $a$ for
$\rho_{c0}=1, B=1, \gamma=.1, H_{c0}=1, c=2, \alpha=.1$.}

\end{figure}

\begin{figure}[!h]
\includegraphics[height=2.5in]{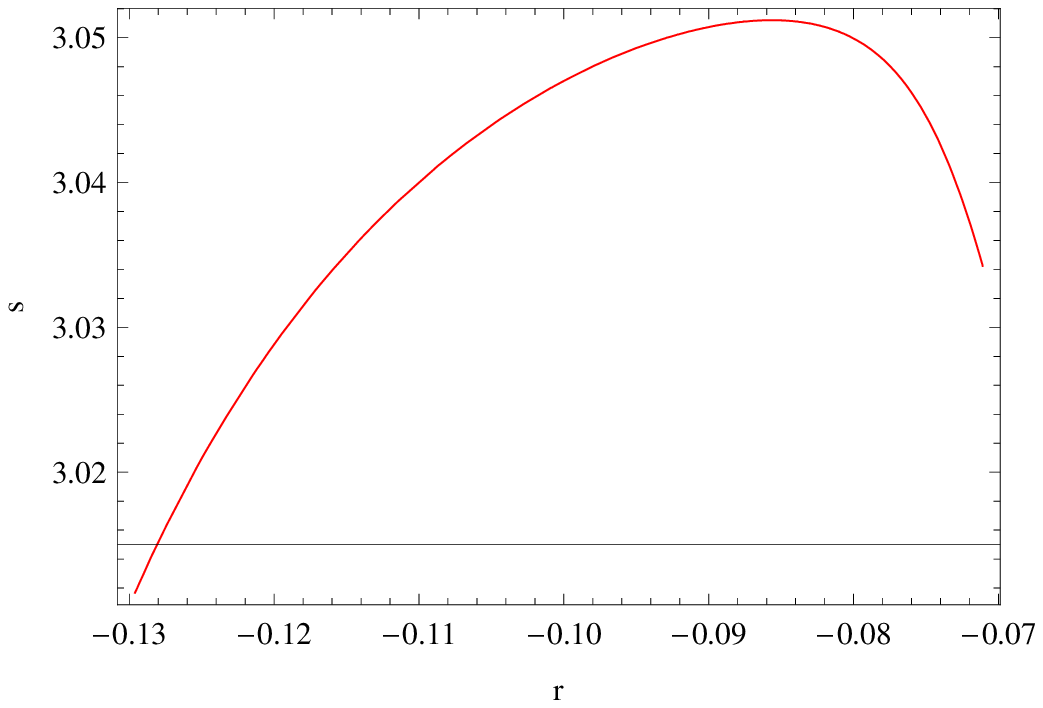}~\\
\vspace{1mm}\caption{The variation of $s$ against $r$ for $
\rho_{c0}=1, H_{c0}=1, c=2, \alpha=.1, B=1, \gamma=.1.$}
\end{figure}

\begin{figure}[!h]
\includegraphics[height=2.5in]{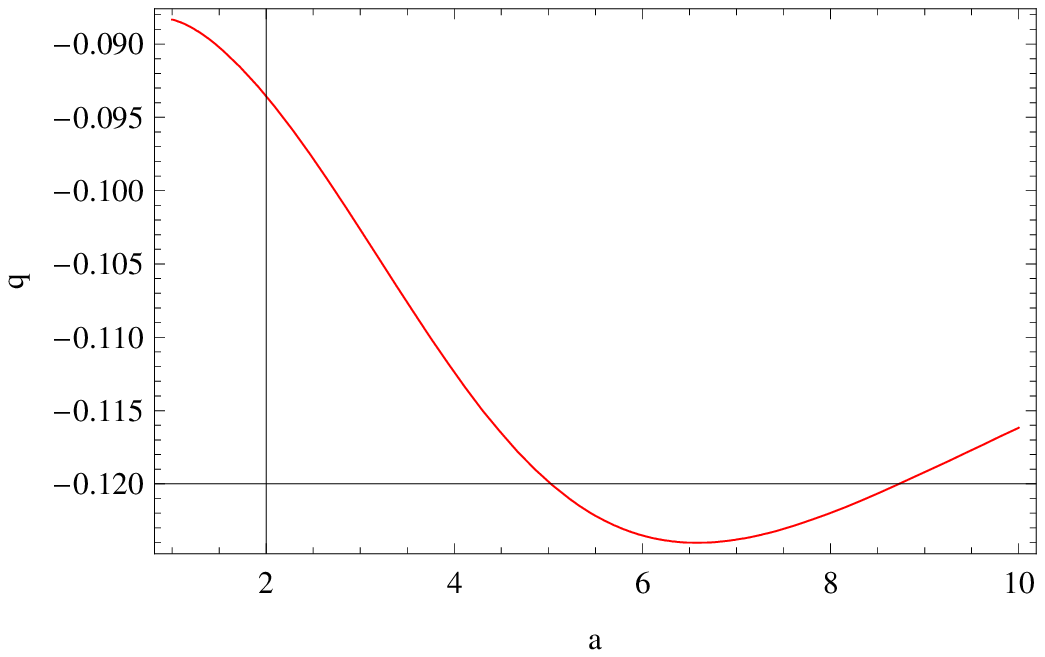}~\\
\vspace{1mm}\caption{The variation of $q$ against $a$ for $
\rho_{c0}=1, H_{c0}=1, c=2, \alpha=.1, B=1, \gamma=.1.$}

\end{figure}

So the equation of state for GHDE $w_{h}$ is obtained as,

\begin{equation}
w_{h}=\frac{p_{h}}{\rho_{h}}=\frac{\frac{(4\alpha c^2 - c^2
+1)}{27\alpha^2 c^4}y(x)I(x)-\frac{1}{9\alpha
c^2}\left[B+\rho_{c0}a^{-3(1+\gamma)}\right]^{\frac{1}{1+\gamma}}+
B\left[B+\rho_{c0}a^{-3(1+\gamma)}\right]^{\frac{-\gamma}{1+\gamma}}+\frac{(4\alpha
c^{2}-c^{2}+1)}{3\alpha c^{2}}H_{c0}^{2}y(x)}{\frac{1}{3\alpha
c^2}y(x)
~I(x)-\left[B+\rho_{c0}a^{-3(1+\gamma)}\right]^{\frac{1}{1+\gamma}}+3H_{c0}^{2}y(x)}
\end{equation}
where, $y(x)=e^{-\frac{(13\alpha c^2-c^2+1)}{3\alpha c^2}x}$.\\

Also the equation state for combined fluid is obtained as,

\begin{equation}
w=\frac{p_{c}+p_{h}}{\rho_{c}+\rho_{h}}=\frac{p_c+\frac{(4\alpha
c^2 - c^2+1)}{27\alpha^2 c^4}y(x)I(x)-\frac{1}{9\alpha
c^2}\left[B+\rho_{c0}a^{-3(1+\gamma)}\right]^{\frac{1}{1+\gamma}}+
B\left[B+\rho_{c0}a^{-3(1+\gamma)}\right]^{\frac{-\gamma}{1+\gamma}}+\frac{(4\alpha
c^{2}-c^{2}+1)}{3\alpha
c^{2}}H_{c0}^{2}y(x)}{\rho_{c}+\frac{1}{3\alpha c^2}y(x)
~I(x)-\left[B+\rho_{c0}a^{-3(1+\gamma)}\right]^{\frac{1}{1+\gamma}}+3H_{c0}^{2}y(x)}
\end{equation}

The above expressions are very completed, so the equation of state
parameters $w_{h}$ and $w$ against $a$ have been drawn in figure
12 and 13 respectively for  $\rho_{c0}=1, H_{c0}=1, c=2,
\alpha=.1, B=1, \gamma=.1$. Also from (8), (9), (10), (46), (50)
and (53) we draw the graphs of $s$ against $r$ and $q$ against $a$
in figures 14 and 15 respectively for  $\rho_{c0}=1, H_{c0}=1,
c=2, \alpha=.1, B=1, \gamma=.1$. We have seen that $s$ first
increases and then decreases as $r$ increases and we see that $s$
is always positive but $r$ keeps negative sign. We have seen that
the EOS for GHDE $w_{h}$ keeps negative sign as evolution of the
universe. When we have considered GHDE and GCG without
interaction, the EOS for combined fluid generates the negative
sign. The deceleration parameter also gives us the negative sign,
so the non-interacting model also generates dark energy. When
$(1-\alpha)$ is replaced by $\beta$, we get the similar results
for GRDE and GCG without interaction.\\

\section{\normalsize\bf{GHDE and GRDE models with GCG : Interacting Scenario}}

Here we consider the universe is filled with the mixture of GCG
and GHDE and also there is an interaction between them. So GCG and
GHDE are not separately conserved and let the interaction term is
defined by $3\delta\hat{} H\rho_{c}$, where $\delta$ is the
interaction parameter. So we get the relations,

\begin{equation}
\dot\rho_{c}+3H(\rho_{c}+p_{c})=-3\delta H\rho_{c}
\end{equation}

and

\begin{equation}
\dot\rho_{h}+3H(\rho_{h}+p_{h})=3\delta H\rho_{c}
\end{equation}

Assume that the equation of state for GCG is
$p_{c}=-B/{\rho}_{c}^{\gamma}$. Putting it in the above equation
and after solving the differential equation we get,

\begin{equation}
\rho_{c}=\left[\frac{B}{1+\delta}+\rho_{c1}
a^{-3(1+\gamma)(1+\delta)}\right]^{\frac{1}{(1+\gamma)}}~~and~~~
p_{c}=-B\left[\frac{B}{1+\delta}+\rho_{c1}
a^{-3(1+\gamma)(1+\delta)}\right]^{\frac{-\gamma}{(1+\gamma)}}
\end{equation}

where $\rho_{c1}$ is the integrating constant. So Thus we get,

\begin{equation}
6\alpha c^{2}\dot{H}+(13\alpha
c^{2}-c^{2}+1)H^2=\frac{1}{3}\left[\frac{B}{1+\delta}+\rho_{c1}
a^{-3(1+\gamma)(1+\delta)}\right]^{\frac{1}{(1+\gamma)}}
\end{equation}

and after solving we get,

\begin{equation}
H^2=\frac{1}{9\alpha c^2}e^{-\frac{(13\alpha c^2-c^2+1)}{3\alpha
c^2}x}\int\left[\frac{B}{1+\delta}+\rho_{c1}
e^{-3(1+\gamma)(1+\delta)x}\right]^{\frac{1}{1+\gamma}}e^{\frac{(13\alpha
c^2-c^2+1)}{3\alpha c^2}x}dx+H_{c1}^{2}a^{-\frac{(13\alpha
c^{2}-c^{2}+1)}{3\alpha c^{2}}}
\end{equation}

where $x=lna$ and $H_{c1}$ is the integrating constant. Hence we
have,

\begin{equation}
\rho_{h}=\frac{1}{3\alpha c^2}e^{-\frac{(13\alpha
c^2-c^2+1)}{3\alpha c^2}x}
~J(x)-\left[\frac{B}{1+\delta}+\rho_{c1}
a^{-3(1+\gamma)(1+\delta)}\right]^{\frac{1}{1+\gamma}}+3H_{c1}^{2}a^{-\frac{(13\alpha
c^{2}-c^{2}+1)}{3\alpha c^{2}}}
\end{equation}
where,
\begin{equation}
J(x)=\int\left[\frac{B}{1+\delta}+\rho_{c1}
e^{-3(1+\gamma)(1+\delta)x}\right]^{\frac{1}{1+\gamma}}e^{\frac{(13\alpha
c^2-c^2+1)}{3\alpha c^2}x}dx
\end{equation}

Now, from energy conservation equation we get,
\begin{equation}
p_{h}=\delta \rho_{c}
-\rho_{h}-\frac{1}{3}.\frac{\partial\rho_{h}}{\partial x}
\end{equation}

which gives,
\begin{eqnarray*}
p_{h}=\frac{(4\alpha c^2 - c^2+1)}{27\alpha^2
c^4}y(x)J(x)+(\delta-\frac{1}{9\alpha
c^2})\left[\frac{B}{1+\delta}+\rho_{c1}
a^{-3(1+\gamma)(1+\delta)}\right]^{\frac{1}{1+\gamma}}
\end{eqnarray*}
\begin{equation}
+ \left[\frac{B}{1+\delta}-\delta\rho_{c1}
a^{-3(1+\gamma)(1+\delta)}\right]\left[\frac{B}{1+\delta}+\rho_{c1}
a^{-3(1+\gamma)(1+\delta)}\right]^{\frac{-\gamma}{1+\gamma}}+\frac{(4\alpha
c^{2}-c^{2}+1)}{3\alpha c^{2}}H_{c1}^{2}y(x)
\end{equation}

So the equation of state for GHDE $w_{h}$ is obtained as,\\

$w_{h}=\frac{p_{h}}{\rho_{h}}=$
 {\tiny{\begin{equation}
=\frac{\frac{(4\alpha c^2 - c^2+1)}{3\alpha^2
c^2}y(x)(\frac{J(x)}{3\alpha
c^2}+H_{c1}^{2})+(\delta-\frac{1}{9\alpha
c^2})[\frac{B}{1+\delta}+\rho_{c1}
a^{-3(1+\gamma)(1+\delta)}]^{\frac{1}{1+\gamma}}+
[\frac{B}{1+\delta}-\delta\rho_{c1}
a^{-3(1+\gamma)(1+\delta)}][\frac{B}{1+\delta}+\rho_{c1}
a^{-3(1+\gamma)(1+\delta)}]^{\frac{-\gamma}{1+\gamma}}}{\frac{1}{3\alpha
c^2}y(x) ~J(x)-\left[\frac{B}{1+\delta}+\rho_{c1}
a^{-3(1+\gamma)(1+\delta)}\right]^{\frac{1}{1+\gamma}}+3H_{c1}^{2}y(x)}
\end{equation}
}}

where, $y(x)=e^{-\frac{(13\alpha c^2-c^2+1)}{3\alpha c^2}x}$.\\

Also the equation state for combined fluid is obtained as,

\begin{equation}
 w=\frac{p_{c}+p_{h}}{\rho_{c}+\rho_{h}}=\frac{\frac{(4\alpha c^2 - c^2+1)}{3\alpha^2
c^2}y(x)(\frac{J(x)}{3\alpha c^2}+H_{c1}^{2})-\frac{1}{9\alpha
c^2}[\frac{B}{1+\delta}+\rho_{c1}
a^{-3(1+\gamma)(1+\delta)}]^{\frac{1}{1+\gamma}}}{(\frac{J(x)}{3\alpha
c^2}+3H_{c1}^{2})y(x)}
\end{equation}

\begin{figure}
\includegraphics[height=2.7in]{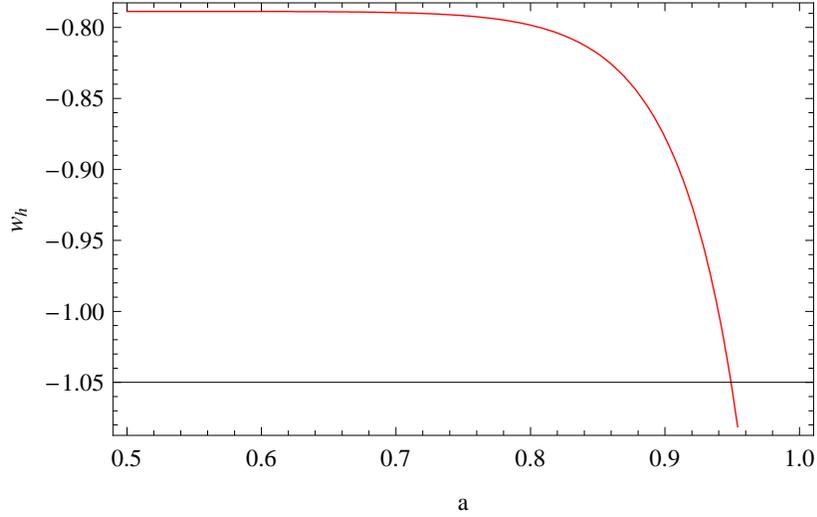}~\\
\vspace{1mm}\caption{The variation of $w_{h}$ against $a$ for
$\gamma=.1, \rho_{c1}=1, H_{c1}=1, c=2, \delta=.01, B=1,
\alpha=.1$}
\end{figure}

\begin{figure}
\includegraphics[height=2.7in]{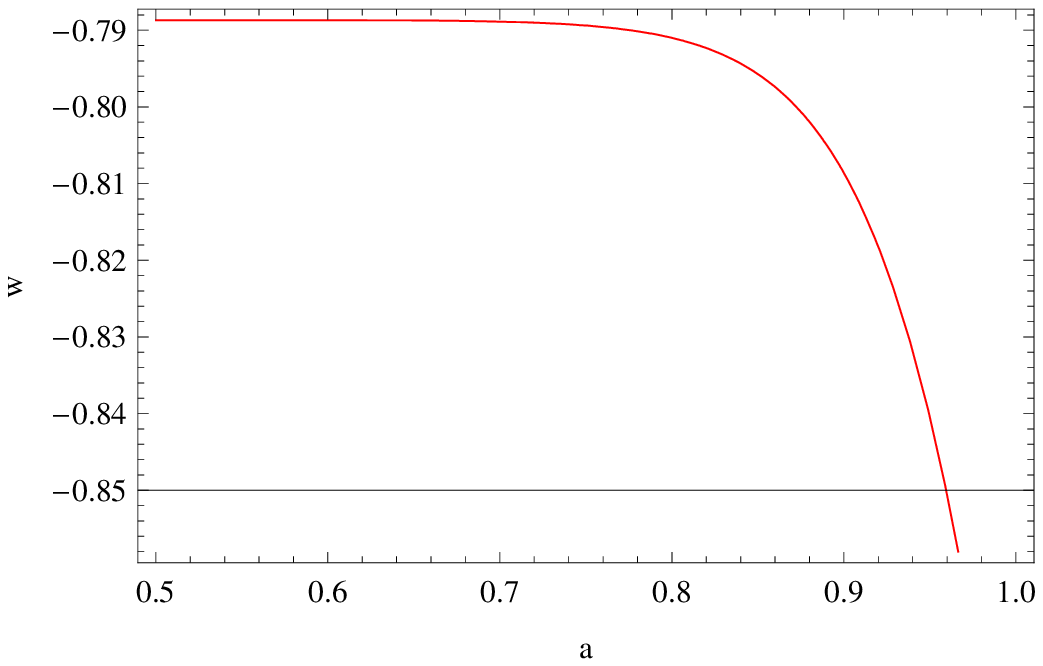}~\\
\vspace{1mm}\caption{The variation of $w$ against $a$ for
$H_{c1}=1, c=2, \gamma=.1, B=1,\alpha=.1, \delta=.01$.}
\end{figure}

\begin{figure}[!h]
\includegraphics[height=2.0in]{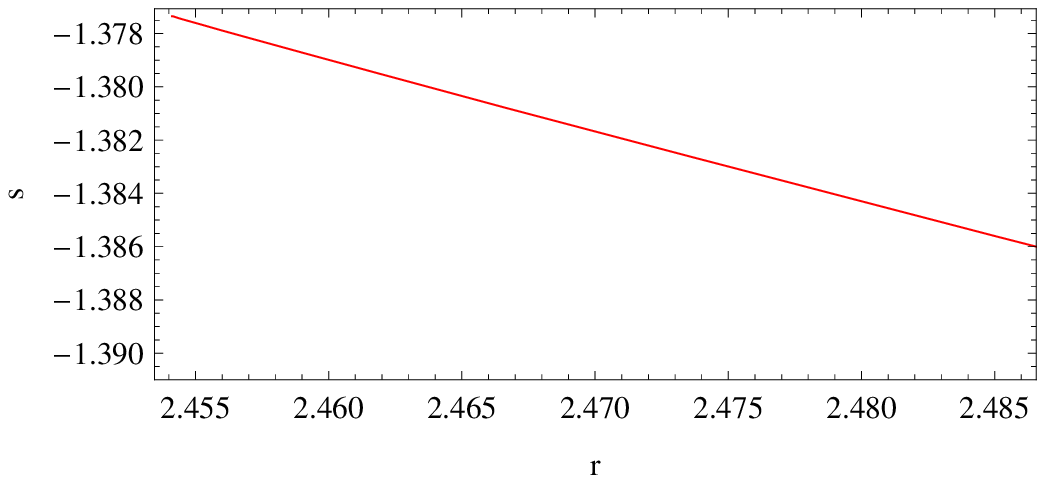}~\\
\vspace{1mm}\caption{The variation of $s$ against $r$ for
$H_{c1}=1, c=2, \gamma=.1, B=1,\alpha=.1, \delta=.01$.}
\end{figure}

\begin{figure}[!h]
\includegraphics[height=2.5in]{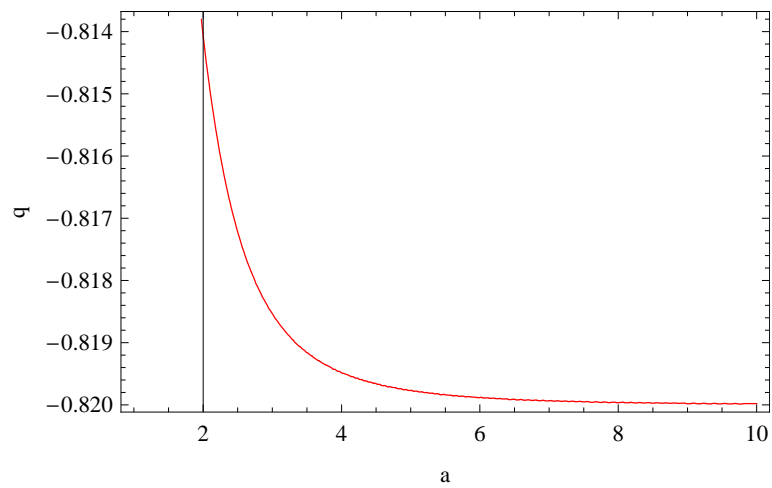}~\\
\vspace{1mm}\caption{The variation of $q$ against $a$ for
$H_{c1}=1, c=2, \gamma=.1, B=1,\alpha=.1, \delta=.01$.}

\end{figure}

The above expressions are very completed, so the equation of state
parameters $w_{h}$ and $w$ against $a$ have been drawn in figure
16 and 17 respectively for $H_{c1}=1, c=2, \gamma=.1,
B=1,\alpha=.01, \delta=.1$. From (8), (9), (10), (46), (50) and
(53), we also draw the graphs of $s$ against $r$ and $q$ against
$a$ in figures 18 and 19 respectively for $H_{c1}=1, c=2,
\gamma=.1, B=1,\alpha=.01, \delta=.1$. We have seen that $s$
always decreases from negative level as $r$ increases and keeps
positive signs. We have seen that the EOS for GHDE $w_{h}$ keeps
negative sign as evolution of the universe. When we have
considered GHDE and GCG with interaction, the EOS for combined
fluid generates the negative sign. The deceleration parameter also
gives us the negative sign, so the interacting model also
generates dark energy. When $\alpha$ is replaced by $(1-\beta)$,
we get the similar results for GRDE and CGC with interaction.\\

\section{\normalsize\bf{Discussions}}

In this work, we have considered the generalized holographic and
generalized Ricci dark energy models for acceleration of the
universe. If the universe filled with only GHDE and GRDE the
corresponding deceleration parameter, EOS parameter and
statefinder parameters have been calculated. The universe will be
accelerating if the parameter $\alpha$ satisfied
$\alpha<\frac{c^{2}-1}{7c^{2}}$ for GHDE model. If the universe is
filled with GRDE rather than GHDE then all the above solutions are
valid provided $\alpha=1-\beta$. So the universe will be
accelerating if $\beta>\frac{8c^{2}-1}{7c^{2}}$. If the universe
is filled with the mixture of dark matter and GHDE/GRDE, the
equation of state parameters and deceleration obey the negative
sign in both interacting and non-interacting scenarios, which
derive the acceleration of the universe. When we consider the
mixture of generalized Chaplygin gas and GHDE/GRDE, the equation
of state parameters and the deceleration parameter also generate
negative sign which shows that the combined fluids derive the
acceleration of the universe. In all the cases, we have verified
the results in drawing the figures of $w_{h}$, $w$ and $q$ by
choosing (in particular) the values $w_{m}=0.01, c=2, \alpha=0.1$
and $\delta=0.01$. The statefinder parameters have different
natures in our interacting and non-interacting cases. If the
universe filled with only GHDE/GRDE, then from the figure 2, we
have seen that when $r$ increases, $s$ first decreases from
positive to negative upto about $r=0.5$ and then $s$ increases
from negative to positive. For non-interacting models of GHDE, we
have seen from figure 6 that, $s$ first increases from some
positive value to $+\infty$ and after that $s$ also increases from
$-\infty$ to some positive value as $r$ increases from negative to
positive and from figure 13, we have also seen that $s$ first
increases and then decreases as $r$ increases and we see that $s$
is always positive but $r$ keeps negative sign. But for
interacting models of GHDE, we have seen from figure 10 that, $s$
always decreases with positive sign as $r$ increases with negative
level and from figure 18 that, $s$ always decreases from negative
level as $r$ increases
and keeps positive signs.\\\\

{\bf Acknowledgement:}\\

The authors are thankful to IUCAA, Pune, India for warm
hospitality where part of the work was carried out.\\\\

{\bf References:}\\\\
$[1]$ A. G. Riess et al, {\it Astron. J.} {\bf 116} 1009 (1998)\\
$[2]$ S. Perlmutter et al, {\it Astrophys. J.} {\bf 517} 565 (1999).\\
$[3]$ N. A. Bachall, J. P. Ostriker, S. Perlmutter and P. J. Steinhardt,
{\it Science} {\bf 284} 1481 (1999).\\
$[4]$ I. Zlatev, L. Wang and P. J. Steinhardt, {\it Phys. Rev.
Lett.} {\bf 82} 896 (1999).\\
$[5]$ V. Sahni, {\it Class. Quantum Grav.} {\bf 19} 3435 (2002).\\
$[6]$ R. R. Caldwell, M. Kamionkowski and N. N. Weinberg, {\it
Phys. Rev. Lett.} {\bf 91} 071301 (2003).\\
$[7]$ B. Feng et al, {\it Phys. Lett. B} {\bf 607} 35 (2005).\\
$[8]$ E. Witten, {\it hep-ph}/0002297.\\
$[9]$ L. Susskind, {\it J. Math. Phys.} {\bf 36} 6377 (1995).\\
$[10]$ A. G. Cohen, D. B. Kaplan and A. E. Nelson, {\it Phys. Rev.
Lett.} {\bf 82} 4971 (1999).\\
$[11]$ X. Zhang, {\it Int. J. Mod. Phys. D} {\bf 14} 1597 (2005).\\
$[12]$ M. R. Setare, {\it Phys. Lett. B} {\bf 648} 329 (2007).\\
$[13]$ W. Fischler and L. Susskind, {\it hep-th}/9806039.\\
$[14]$ Y. Gong, {\it Phys. Rev. D} {\bf 70} 064029 (2004).\\
$[15]$ M. Li, {\it Phys. Lett. B} {\bf 603} 1 (2004).\\
$[16]$ C. Gao, F. Wu, X. Chen and Y. -G. Shen, {\it Phys. Rev. D}
{\bf 79} 043511 (2009).\\
$[17]$ R. G. Cai, B. Hu and Y. Zhang, {\it Commun. Theor. Phys.} {\bf 51} 954 (2009).\\
$[18]$ C. -J. Feng, {\it Phys. Lett. B} {\bf 670} 231 (2008); {\it
Phys. Lett. B} {\bf 672} 94 (2009); C. -J. Feng and X. -Z. Li,
{\it Phys. Lett. B} {\bf 680} 355 (2009); L. Xu, W. Li and J. Lu,
{\it Mod. Phys. Lett. A} {\bf 24} 1355
(2009); M. Suwa and T. Nihei, {\it Phys. Rev. D} {\bf 81} 023519 (2010).\\
$[19]$ L. Xu, J. Lu and W. Li., {\it Eur. Phys. J. C} {\bf 64} 89 (2009).\\
$[20]$ V. Sahni, T. D. Saini, A. A. Starobinsky and U. Alam, {\it JETP Lett.} {\bf 77} 201 (2003).\\
$[21]$ K. Y. Kim, H. W. Lee and Y. S. Myung, {\it Gen. Rel. Grav.} {\bf 43} 1095 (2011).\\
$[22]$ V. Gorini, A. Kamenshchik and U. Moschella, {\it Phys. Rev.
D} {\bf 67} 063509 (2003); U. Alam, V. Sahni , T. D. Saini and
A.A. Starobinsky, {\it Mon. Not. Roy. Astron. Soc.} {\bf 344}, 1057 (2003).\\

\end{document}